\documentclass[preprint,showpacs,preprintnumbers,amsmath,amssymb]{revtex4}

\usepackage{graphicx}
\usepackage{dcolumn}
\usepackage{bm}
\usepackage{hyperref}
\hypersetup{colorlinks=false}

\begin{document}

\newcommand{\RR}[1]{[#1]}
\newcommand{\intsum}{\sum \kern -15pt \int}
\newfont{\Yfont}{cmti10 scaled 2074}
\newcommand{\Y}{\hbox{{\Yfont y}\phantom.}}
\def\O{{\cal O}}
\newcommand{\bra}[1]{\left< #1 \right| }
\newcommand{\braa}[1]{\left. \left< #1 \right| \right| }
\def\Bra#1#2{{\mbox{\vphantom{$\left< #2 \right|$}}}_{#1}
\kern -2.5pt \left< #2 \right| }
\def\Braa#1#2{{\mbox{\vphantom{$\left< #2 \right|$}}}_{#1}
\kern -2.5pt \left. \left< #2 \right| \right| }
\newcommand{\ket}[1]{\left| #1 \right> }
\newcommand{\kett}[1]{\left| \left| #1 \right> \right.}
\newcommand{\scal}[2]{\left< #1 \left| \mbox{\vphantom{$\left< #1 #2 \right|$}}
\right. #2 \right> }
\def\Scal#1#2#3{{\mbox{\vphantom{$\left<#2#3\right|$}}}_{#1}
{\left< #2 \left| \mbox{\vphantom{$\left<#2#3\right|$}} \right. #3
\right> }}


\title{A New Expression for 3N Bound State Faddeev Equation\\in a 3D Approach}

\author{M. Harzchi}
\email{mehdi\underline{ }harzchi@khayam.ut.ac.ir}
\author{S. Bayegan}%
\email{bayegan@khayam.ut.ac.ir}

\affiliation{Department of Physics, University of Tehran, P.O.Box
 14395-547, Tehran, Iran. }%

\date{\today}

\begin{abstract}
A spin-isospin dependent three-dimensional approach has been applied
for formulation of the three-nucleon bound state and a new
expression for Faddeev equation based on three-nucleon free basis
state has been obtained. Then the three-nucleon wave function has
been obtained as a function of five independent variables.
\end{abstract}

\maketitle

\section{Introduction}
\label{intro} \label{sec:introduction} During the past years, the
three-dimensional (3D) approach has been developed for few-body
bound and scattering problems~\cite{WSchadow}-\cite{Harzchi}. The
motivation for developing this approach is introducing a direct
solution of the integral equations avoiding the very involved
angular momentum algebra occurring for the permutations,
transformations and especially for the three-body forces.

In the case of the three-body bound state the Faddeev equation has
been formulated for three identical bosons as a function of vector
Jacobi momenta, with the specific stress upon the magnitudes of the
momenta and the angles between them~\cite{Schadow}. Adding the
spin-isospin to the 3D formalism was a major additional task which
carried out in Ref.~\cite{Bayegan}. In this paper we have attempted
to reformulate the three-nucleon (3N) bound state and have obtained
a new expression for Faddeev integral equation. To this end we have
used 3N free basis state for representation of 3N wave function.

This manuscript is organized as follow. In Sect.~\ref{sec:Faddeev}
we have derived a new expression for Faddeev equation in a realistic
3D scheme as a function of Jacobi momenta vectors and the
spin-isospin quantum numbers. Then we have chosen suitable
coordinate system for describing Faddeev component of total 3N wave
function as function of five independent variables for numerical
calculations. Finally in Sect.~\ref{sec:summary} a summary and an
outlook have been presented.

\section{3N bound state in a 3D momentum representation}\label{sec:Faddeev}
\subsection{Faddeev equation}
\label{sec:2} Faddeev equation for the 3N bound state with
considering pairwise-interactions is described by~\cite{Stadler}:
\begin{eqnarray}\label{eq.1}
|\psi^{M_{t}}\rangle=G_{0}tP|\psi^{M_{t}}\rangle,
\end{eqnarray}
where $|\psi^{M_{t}}\rangle$ is Faddeev component of the total 3N
wave function, $M_{t}$ bing the projection of total angular momentum
along the quantization axis, $P=P_{12}P_{23}+P_{13}P_{23}$ is the
sum of cyclic and anti-cyclic permutations of three nucleons, $t$
denotes the two-body transition operator which is determined by a
Lippmann-Schwinger equation and $G_{0}$ is the free 3N propagator
which is given by:
\begin{eqnarray}\label{eq.2}
G_{0}=\frac{1}{E-\frac{p^{2}}{m}-\frac{3q^{2}}{4m}},
\end{eqnarray}
where $E$ is the binding energy of 3N bound state. In order to solve
Eq.~(\ref{eq.1}) in the momentum space we introduce the 3N free
basis state in a 3D formalism as~\cite{Fachruddin-PRC68}:
\begin{eqnarray}\label{eq.3}
|\mathbf{p}\mathbf{q}\gamma\rangle\equiv|\mathbf{p}\mathbf{q}\,m_{s_{1}}
m_{s_{2}}m_{s_{3}}m_{t_{1}}m_{t_{2}}m_{t_{3}}\rangle\equiv|\mathbf{q}
\,m_{s_{1}}m_{t_{1}}\rangle|\mathbf{p}\,m_{s_{2}}m_{s_{3}}m_{t_{2}}m_{t_{3}}\rangle,
\end{eqnarray}
This basis state involves two standard Jacobi momenta $\mathbf{p}$
and $\mathbf{q}$ which are the relative momentum vector in the
subsystem and the momentum vector of the spectator with respect to
the subsystem respectively~\cite{Stadler}.
$|\gamma\rangle\equiv|m_{s_{1}}m_{s_{2}}m_{s_{3}}m_{t_{1}}m_{t_{2}}m_{t_{3}}\rangle$
is the spin-isospin parts of the basis state where the quantities
$m_{s_{i}}$($m_{t_{i}}$) are the projections of the spin (isospin)
of each three nucleons along the quantization axis. The introduced
basis states are completed and normalized as:
\begin{eqnarray}\label{eq.4}
\sum_{\gamma}\int d\mathbf{p}\int
d\mathbf{q}\,|\mathbf{p}\mathbf{q}\gamma\rangle\langle\mathbf{p}\mathbf{q}\gamma|=
1,\quad\quad\quad\;
\langle\mathbf{p}'\mathbf{q}'\gamma'|\mathbf{p}\mathbf{q}\gamma\rangle=
\delta(\mathbf{p}'-\mathbf{p})\,\delta(\mathbf{q}'-\mathbf{q})\,\delta_{\gamma'\gamma}.
\end{eqnarray}
Now we start by inserting the completeness relation twice into
Eq.~(\ref{eq.1}) as follow:
\begin{eqnarray}\label{eq.6}
\langle\mathbf{p}\mathbf{q}\gamma|\psi^{M_{t}}
\rangle&=&\frac{1}{E-\frac{p^{2}}{m}-\frac{3q^{2}}{4m}}\sum_{\gamma''}\int
d\mathbf{p}''\int d\mathbf{q}''\sum_{\gamma'}\int d\mathbf{p}'\int
d\mathbf{q}'\nonumber\\&&\times\,\langle\mathbf{p}\mathbf{q}\gamma|t|\mathbf{p}''\mathbf{q}''\gamma''
\rangle\langle\mathbf{p}''\mathbf{q}''\gamma''|P|\mathbf{p}'\mathbf{q}'\gamma'\rangle
\langle\mathbf{p}'\mathbf{q}'\gamma'|\psi^{M_{t}}\rangle.
\end{eqnarray}
The matrix elements of the permutation operator $P$ are evaluated
as~\cite{Fachruddin-PRC68}:
\begin{eqnarray} \label{eq.7}
&&\langle\mathbf{p}''\mathbf{q}''\gamma''|P|\mathbf{p}'\mathbf{q}'\gamma'\rangle\nonumber\\&=&\delta(\mathbf{p}''-\frac{1}{2}\mathbf{q}''-\mathbf{q}')\,\delta(\mathbf{p}'+\mathbf{q}''+\frac{1}{2}
\mathbf{q}')\,\delta_{m''_{s_{1}}m'_{s_{3}}}\delta_{m''_{s_{2}}m'_{s_{1}}}
\delta_{m''_{s_{3}}m'_{s_{2}}}\delta_{m''_{t_{1}}m'_{t_{3}}}\delta_{m''_{t_{2}}m'_{t_{1}}}\delta_{m''_{t_{3}}
m'_{t_{2}}}\nonumber\\&&+\,\delta\,(\mathbf{p}''+\frac{1}{2}\mathbf{q}''+\mathbf{q}')
\,\delta(\mathbf{p}'-\mathbf{q}''-\frac{1}{2}\mathbf{q}')\,\delta_{m''_{s_{1}}
m'_{s_{2}}}\delta_{m''_{s_{2}}m'_{s_{3}}}\delta_{m''_{s_{3}}m'_{s_{1}}}\delta_{m''_{t_{1}}m'_{t_{2}}}
\delta_{m''_{t_{2}}m'_{t_{3}}}\delta_{m''_{t_{3}}m'_{t_{1}}},\quad\quad
\end{eqnarray}
and for the two-body $t$-matrix we have:
\begin{eqnarray}\label{eq.8}
&&\langle\mathbf{p}\mathbf{q}\gamma|t|\mathbf{p}''\mathbf{q}''\gamma''\rangle=\langle\mathbf{p}\,m_{s_{2}}m_{s_{3}}m_{t_{2}}m_{t_{3}}|t(\epsilon)|\mathbf{p}''m''_{s_{2}}m''_{s_{3}}
m''_{t_{2}}m''_{t_{3}}\rangle\,\delta(\mathbf{q}-\mathbf{q}'')\,\delta_{m_{s_{1}}m''_{s_{1}}}
\,\delta_{m_{t_{1}}m''_{t_{1}}},\quad
\end{eqnarray}
where $\epsilon=E-\frac{3}{4m}q^{2}$, is the energy carried by a
two-body subsystem in a three-nucleon system. Substituting
Eqs.~(\ref{eq.7}) and (\ref{eq.8}) into Eq.~(\ref{eq.6}) yields:
\begin{eqnarray}\label{eq.9}
&&\langle\mathbf{p}\mathbf{q}\gamma|\,\psi^{M_{t}}\rangle\nonumber\\&=&
\frac{1}{E-\frac{{p}^{2}}{m}-\frac{3q^{2}}{4m}}\sum_{m'_{s_{1}}m'_{t_{1}}}\int
d\mathbf{q}'\nonumber\\&&\times\Bigl\{\sum_{m'_{s_{2}}m'_{t_{2}}}\langle\mathbf{p}\,m_{s_{2}}m_{s_{3}}m_{t_{2}}m_{t_{3}}|t(\epsilon)|\boldsymbol\pi\,
m'_{s_{1}}m'_{s_{2}}m'_{t_{1}}m'_{t_{2}}\rangle\langle-\boldsymbol\pi'\mathbf{q}'m'_{s_{1}}m'_{s_{2}}m_{s_{1}}m'_{t_{1}}m'_{t_{2}}m_{t_{1}}|\psi^{M_{t}}\rangle\nonumber
\\&&+\sum_{m'_{s_{3}}m'_{t_{3}}}\langle\mathbf{p}\,m_{s_{2}}m_{s_{3}}m_{t_{2}}m_{t_{3}}|t(\epsilon)|-\boldsymbol\pi\,m'_{s_{3}}m'_{s_{1}}m'_{t_{3}}m'_{t_{1}}\rangle\langle\boldsymbol\pi'\mathbf{q}'m'_{s_{1}}m_{s_{1}}m'_{s_{3}}m'_{t_{1}}m_{t_{1}}m'_{t_{3}}|\psi^{M_{t}}\rangle\Bigl\}\nonumber
\nonumber\\&=&\frac{1}{E-\frac{{p}^{2}}{m}-\frac{3q^{2}}{4m}}\sum_{m'_{s_{1}}m'_{t_{1}}m'_{s}m'_{t}}\int
d\mathbf{q}'\nonumber
\\&&\times\Bigl \{\langle\mathbf{p}\,m_{s_{2}}m_{s_{3}}m_{t_{2}}m_{t_{3}}|t(\epsilon)|\boldsymbol\pi\,m'_{s_{1}}m'_{s}m'_{t_{1}}m'_{t}\rangle\langle\boldsymbol\pi'\mathbf{q}'m'_{s_{1}}m_{s_{1}}m'_{s}m'_{t_{1}}m'_{t}m_{t_{1}}|P_{23}|\psi^{M_{t}}\rangle\nonumber
\\&&+\langle\mathbf{p}\,m_{s_{2}}m_{s_{3}}m_{t_{2}}m_{t_{3}}|t(\epsilon)P_{23}|\boldsymbol\pi\,m'_{s_{1}}m'_{s}m'_{t_{1}}m'_{t}\rangle\langle\boldsymbol\pi'\mathbf{q}'m'_{s_{1}}m_{s_{1}}m'_{s}m'_{t_{1}}m_{t_{1}}m'_{t}|\psi^{M_{t}}\rangle
\Bigl \}.
\end{eqnarray}\label{eq.10}In the last equality we have used the antisymmetry of Faddeev
component of the 3N wave function as:
\begin{eqnarray}\label{eq.11}
P_{23}|\psi^{M_{t}}\rangle=-|\psi^{M_{t}}\rangle,
\end{eqnarray}
and also we have considered:
\begin{eqnarray}\label{eq.12}
\boldsymbol\pi=\frac{1}{2}\mathbf{q}+\mathbf{q}',\quad\boldsymbol\pi'=\mathbf{q}+\frac{1}{2}\mathbf{q}'.
\end{eqnarray}
The antisymmetrized two-body $t$-matrix is introduced
as~\cite{IFachruddin}:
\begin{eqnarray}\label{eq.12}
\,_{a}\langle\mathbf{p}'m'_{s_{2}}m'_{s_{3}}m'_{t_{2}}m'_{t_{3}}|t|\mathbf{p}m_{s_{2}}m_{s_{3}}m_{t_{2}}
m_{t_{3}}\rangle_{a}=\langle\mathbf{p}'\,m'_{s_{2}}m'_{s_{3}}m'_{t_{2}}m'_{t_{3}}|t(1-P_{23})|
\mathbf{p}\,m_{s_{2}}m_{s_{3}}m_{t_{2}}m_{t_{3}}\rangle,\quad
\end{eqnarray}
where $|\mathbf{p}\,m_{s_{2}}m_{s_{3}}m_{t_{2}}m_{t_{3}}\rangle_{a}$
is the antisymmetrized two-body state which is defined as:
\begin{eqnarray}\label{eq.13}
|\mathbf{p}\,m_{s_{2}}m_{s_{3}}m_{t_{2}}m_{t_{3}}\rangle_{a}=\frac{1}{\sqrt{2}}(1-P_{23})|\mathbf{p}
\,m_{s_{2}}m_{s_{3}}m_{t_{2}}m_{t_{3}}\rangle.
\end{eqnarray}
Hence the final expression for Faddeev equation is explicitly
written:
\begin{eqnarray}\label{eq.14}
\langle\mathbf{p}\mathbf{q}\gamma|\psi^{M_{t}}\rangle&=&\frac{-1}{E-\frac{{p}^{2}}{m}-\frac{3q^{2}}{4m}}\sum_{m'_{s_{1}}m'_{t_{1}}m'_{s}m'_{t}}\int
d\mathbf{q}'\,_{a}\langle\mathbf{p}\,m_{s_{2}}m_{s_{3}}m_{t_{2}}m_{t_{3}}|t(\epsilon)|\boldsymbol\pi\,m'_{s_{1}}m'_{s}m'_{t_{1}}m'_{t}\rangle_{a}\nonumber
\\&&\times\langle\boldsymbol\pi'\mathbf{q}'m'_{s_{1}}m_{s_{1}}m'_{s}m'_{t_{1}}m_{t_{1}}m'_{t}|\psi^{M_{t}}\rangle.\quad\quad
\end{eqnarray}
As a simplification we rewrite this equation as:
\begin{eqnarray}\label{eq.15}
\psi^{M_{t}}_{\gamma}(\mathbf{p},\mathbf{q})&=&\frac{-1}{E-\frac{{p}^{2}}{m}-\frac{3q^{2}}{4m}}\sum_{m''_{s}m''_{t}m'_{s}m'_{t}}\int
d\mathbf{q}'
\,t_{a}\,_{m_{s_{2}}m_{s_{3}}m_{t_{2}}m_{t_{3}}}^{m''_{s}m'_{s}m''_{t}m'_{t}}(\mathbf{p},\boldsymbol\pi;\epsilon)\,\psi^{M_{t}}_{\tilde{\gamma}}(\boldsymbol\pi',\mathbf{q}'),
\end{eqnarray}
where we have used index $\tilde{\gamma}$ instead of
$m''_{s}m_{s_{1}}m'_{s}m''_{t}m_{t_{1}}m'_{t}$ for simplicity. This
new expression is more simple for numerical calculations in
comparison with previous expression which has been presented in
Ref.~\cite{Bayegan}:
\begin{eqnarray}\label{eq.15}
\psi^{M_{t}}_{\alpha}(\mathbf{p},\mathbf{q})&=&\frac{1}{E-\frac{{p}^{2}}{m}-\frac{3q^{2}}{4m}}\sum_{\gamma\gamma'\alpha'}g_{\alpha\gamma}\,g_{\gamma'\alpha'}\,\delta_{m_{s_{3}}m'_{s_{1}}}\delta_{m_{t_{3}}m'_{t_{1}}}\nonumber\\
&&\times\int d\mathbf{q}'
\,t_{a}\,_{m_{s_{1}}m_{s_{2}}m_{t_{1}}m_{t_{2}}}^{m'_{s_{2}}m'_{s_{3}}m'_{t_{2}}m'_{t_{3}}}(\mathbf{p},-\boldsymbol\pi;\epsilon)\,\psi^{M_{t}}_{\alpha'}(\boldsymbol\pi',\mathbf{q}'),
\end{eqnarray}
For solving the Eq.~(14) one needs the matrix elements of the
antisymmetrized two-body $t$-matrix. We connect this quantity to its
momentum-helicity representation in appendix A. To solve this
integral equation numerically, we have to define a suitable
coordinate system. It is convenient to choose the spin polarization
direction parallel to the $z$ axis and express the momentum vectors
in this coordinate system. With this selection we can write the
two-body $t$-matrix and 3N wave function as (see appendices A and
B):
\begin{eqnarray}\label{eq.16}
&&t_{a}\,_{m_{s_{2}}m_{s_{3}}m_{t_{2}}m_{t_{3}}}^{m''_{s}m'_{s}m''_{t}m'_{t}}(\mathbf{p},\boldsymbol\pi;\epsilon)\nonumber\\&=&e^{-i[(m_{s_{2}}+m_{s_{3}})\varphi_{p}-(m''_{s}+m'_{s})\varphi_{\pi}]}
\,t_{a}^{\pm}\,_{m_{s_{2}}m_{s_{3}}m_{t_{2}}m_{t_{3}}}^{m''_{s}m'_{s}m''_{t}m'_{t}}(p,x_{p},\cos\varphi_{p\pi},x_{\pi},\pi,y_{p\pi};\epsilon),\,\,\,\,\,
\end{eqnarray}
\begin{eqnarray}\label{eq.17}
\psi^{M_{t}}_{\gamma}(\mathbf{p},\mathbf{q})&=&e^{-i[(m_{s_{2}}+m_{s_{3}})\varphi_{p}+(m_{s_{1}}-M_{t})\varphi_{q}]}
\,\,\,^{\pm}\psi^{M_{t}}_{\gamma}(p,x_{p},\cos\varphi_{pq},x_{q},q),\quad\quad
\end{eqnarray}
\begin{eqnarray}\label{eq.18}
\psi^{M_{t}}_{\tilde{\gamma}}(\boldsymbol\pi',\mathbf{q}')
&=&e^{-i[(m_{s_{1}}+m'_{s})\varphi_{\pi'}+(m''_{s}-M_{t})\varphi']}\,\,\,^{\pm}\psi^{M_{t}}_{\tilde{\gamma}}(\pi',x_{\pi'},\cos\varphi_{\pi'q'},x',q').\,\,\,
\end{eqnarray}
where $x'=\hat{\textbf{q}}'\cdot\hat{\textbf{z}}$,
$\varphi'=\varphi_{q'}$ and the labels $\pm$ are related to the
signs of $\sin\varphi_{p\pi}$, $\sin\varphi_{pq}$ and
$\sin\varphi_{\pi'q'}$. With considering:
\begin{eqnarray}\label{eq.19}
&&\varphi_{\pi}=\varphi'+\varphi_{\pi
q'},\,\,\,\quad\,\,\,\varphi_{\pi}'=\varphi'+\varphi_{\pi'q'}.
\end{eqnarray}
Eq.~(14) can be written as:
\begin{eqnarray}\label{eq.20}
&&\,^{\pm}\psi^{M_{t}}_{\gamma}(p,x_{p},\cos\varphi_{pq},x_{q},q)\nonumber\\
&=&-\sum_{m'_{s}m'_{t}m''_{s}m''_{t}}\int_{0}^{\infty}
dq'\int_{-1}^{1}dx'\int_{0}^{2\pi}d\varphi'
\,e^{i(m''_{s}+m'_{s})\varphi_{\pi
q'}}\,e^{-i(m_{s_{1}}+m'_{s})\varphi_{\pi'q'}}\,e^{i(m_{s_{1}}-M_{t})(\varphi_{q}-\varphi')}
\nonumber\\&&\times
\,t_{a}\,_{m_{s_{2}}m_{s_{3}}m_{t_{2}}m_{t_{3}}}^{m''_{s}m'_{s}m''_{t}m'_{t}}(p,x_{p},\cos\varphi_{p\pi},x_{\pi},\pi,y_{p\pi};\epsilon)\,\,\,^{\pm}\psi^{M_{t}}_{\tilde{\gamma}}(\pi',x_{\pi'},\cos\varphi_{\pi'q'},x',q'),
\end{eqnarray}
where the variables are developed similar to the 3N scattering
as~\cite{Harzchi}:
\begin{eqnarray}\label{eq.21}
x_{q}&=&\hat{\textbf{q}}\cdot\hat{\textbf{z}},\nonumber\\x_{p}&=&\hat{\textbf{p}}\cdot\hat{\textbf{z}},\nonumber\\\pi&=&\sqrt{\frac{1}{4}q^{2}+q'^{2}+qq'y_{qq'}},\nonumber\\\pi'&=&\sqrt{q^{2}+\frac{1}{4}q'^{2}+qq'y_{qq'}},\nonumber\\x_{\pi}&=&\hat{\boldsymbol\pi}\cdot\hat{\textbf{z}}=\frac{\frac{1}{2}qx_{q}+q'x'}{\pi},\nonumber\\x_{\pi'}&=&\hat{\boldsymbol\pi}'\cdot\hat{\textbf{z}}=\frac{qx_{q}+\frac{1}{2}q'x'}{\pi'},\nonumber\\y_{p\pi}&=&\hat{\textbf{p}}\cdot\hat{\boldsymbol\pi}=\frac{\frac{1}{2}qy_{pq}+q'y_{pq'}}{\pi},\nonumber\\y_{\pi q'}&=&\hat{\boldsymbol\pi}\cdot\hat{\textbf{q}}'=\frac{\frac{1}{2}qy_{qq'}+q'}{\pi},\nonumber\\
y_{\pi'q'}&=&\hat{\boldsymbol\pi}'\cdot\hat{\textbf{q}}'=\frac{qy_{qq'}+\frac{1}{2}q'}{\pi'},\nonumber\\y_{pq}&=&\hat{\textbf{p}}\cdot\hat{\textbf{q}}=x_{p}x_{q}+\sqrt{1-x_{p}^{2}}\sqrt{1-x_{q}^{2}}\cos\varphi_{pq},\nonumber\\y_{pq'}&=&\hat{\textbf{p}}\cdot\hat{\textbf{q}}'=x_{p}x'+\sqrt{1-x_{p}^{2}}\sqrt{1-x'^{2}}\cos(\varphi_{p}-\varphi'),\nonumber\\y_{qq'}&=&\hat{\textbf{q}}\cdot\hat{\textbf{q}}'=x_{q}x'+\sqrt{1-x_{q}^{2}}\sqrt{1-x'^{2}}\cos(\varphi_{q}-\varphi'),\nonumber\\
\cos\varphi_{p\pi}&=&\frac{\hat{\textbf{p}}\cdot\hat{\boldsymbol\pi}-(\hat{\textbf{p}}\cdot\hat{\textbf{z}})(\hat{\boldsymbol\pi}\cdot\hat{\textbf{z}})}{\sqrt{1-(\hat{\textbf{p}}\cdot\hat{\textbf{z}})^{2}}\sqrt{1-(\hat{\boldsymbol\pi}\cdot\hat{\textbf{z}})^{2}}}=\frac{y_{p\pi}-x_{p}x_{\pi}}{\sqrt{1-x_{p}^{2}}\sqrt{1-x_{\pi}^{2}}},
\nonumber\\
\cos\varphi_{\pi'q'}&=&\frac{\hat{\boldsymbol\pi}'\cdot\textbf{q}'-(\hat{\boldsymbol\pi}'\cdot\hat{\textbf{z}})(\hat{\textbf{q}}'\cdot\hat{\textbf{z}})}{\sqrt{1-(\hat{\boldsymbol\pi}'\cdot\hat{\textbf{z}})^{2}}\sqrt{1-(\hat{\textbf{q}}'\cdot\hat{\textbf{z}})^{2}}}=\frac{y_{\pi'q'}-x_{\pi}'x'}{\sqrt{1-x_{\pi'}^{2}}\sqrt{1-x'^{2}}},
\nonumber\\
\cos\varphi_{\pi
q'}&=&\frac{\hat{\boldsymbol\pi}\cdot\textbf{q}'-(\hat{\boldsymbol\pi}\cdot\hat{\textbf{z}})(\hat{\textbf{q}}'\cdot\hat{\textbf{z}})}{\sqrt{1-(\hat{\boldsymbol\pi}\cdot\hat{\textbf{z}})^{2}}\sqrt{1-(\hat{\textbf{q}}'\cdot\hat{\textbf{z}})^{2}}}=\frac{y_{\pi
q'}-x_{\pi}x'}{\sqrt{1-x_{\pi}^{2}}\sqrt{1-x'^{2}}}.
\end{eqnarray}
It is clear that the Faddeev component of the wave function $\psi$
is explicitly calculated as function of five independent variables.
In appendices D and E we discuss about the $\varphi'$- and
$x'$-integration and also determination of the signs of sine
functions without any ambiguity.

In this stage we discuss about the total number of coupled integral
equations. The total number of coupled Faddeev equations for the 3N
bound state in a realistic 3D formalism according to the
spin-isospin states is given by:
\begin{eqnarray}
N=2\,(N_{t}\times
N_{s})=2\,(N_{t}\times\sum_{i=1}^{3}N_{m_{s_{i}}}),
\end{eqnarray}
where $N_{s}$ and $N_{t}$ are the total number of spin and isospin
states respectively and $N_{m_{s_{i}}}$ is the number of spin states
for each nucleon. It is clear that $N_{m_{s_{i}}}=2$ and $N_{t}=3$
for our problem. The factor 2 is related to signs of sine functions
of azimuthal angles which is explained in appendix D. Consequently
the total number of coupled Faddeev equations for either $\,^{3}$H
and $\,^{3}$He is $N=48$.
\subsection{Total wave function}
The total 3N wave function $|\Psi^{M_{t}}\rangle$ is given
by~\cite{Stadler}:
\begin{eqnarray}\label{eq.22}
|\Psi^{M_{t}}\rangle=(1+P)|\psi^{M_{t}}\rangle.
\end{eqnarray}
Now we derive an expression for the matrix elements of the total 3N
wave function by inserting the 3N free basis state as fallow:
\begin{eqnarray}\label{eq.23}
\langle\mathbf{p}\mathbf{q}\gamma|\Psi^{M_{t}}\rangle
&=&\langle\mathbf{p}\mathbf{q}\gamma|\psi^{M_{t}}\rangle
+\langle\mathbf{p}\mathbf{q}\gamma|P_{12}P_{23}|\psi^{M_{t}}\rangle+\langle\mathbf{p}\mathbf{q}\gamma|P_{13}P_{23}|\psi^{M_{t}}\rangle.
\end{eqnarray}
By applying the permutation operator $P_{12}P_{23}$ and
$P_{13}P_{23}$ to the 3N free basis state, Eq.~(\ref{eq.23}) can be
written as~\cite{Fachruddin-PRC68}:
\begin{eqnarray}\label{eq.24}
\langle\mathbf{p}\mathbf{q}\gamma|\Psi^{M_{t}}\rangle=\langle\mathbf{p}\mathbf{q}\gamma|\psi^{M_{t}}\rangle+\langle\textbf{p}_{2}\textbf{q}_{2}\gamma_{2}|\psi^{M_{t}}\rangle
+\langle\textbf{p}_{3}\textbf{q}_{3}\gamma_{3}|\psi^{M_{t}}\rangle,
\end{eqnarray}
with:
\begin{eqnarray}\label{eq.25}
\textbf{p}_{2}&=&-\frac{1}{2}\mathbf{p}-\frac{3}{4}\mathbf{q},\quad\quad\textbf{q}_{2}=\mathbf{p}-\frac{1}{2}\mathbf{q},\quad\quad\gamma_{2}\equiv
m_{s_{2}}m_{s_{3}}m_{s_{1}}m_{t_{2}}m_{t_{3}}m_{t_{1}},\nonumber\\
\textbf{p}_{3}&=&-\frac{1}{2}\mathbf{p}+\frac{3}{4}\mathbf{q},\quad\quad\textbf{q}_{3}=-\mathbf{p}-\frac{1}{2}\mathbf{q},\quad\,\gamma_{3}\equiv
m_{s_{3}}m_{s_{1}}m_{s_{2}}m_{t_{3}}m_{t_{1}}m_{t_{2}}.
\end{eqnarray}
As a simplification Eq.~(\ref{eq.24}) is rewritten as:
\begin{eqnarray}\label{eq.26}
\Psi^{M_{t}}_{\gamma}(\mathbf{p},\mathbf{q})=\psi^{M_{t}}_{\gamma}(\mathbf{p},\mathbf{q})+\psi^{M_{t}}_{\gamma_{2}}(\mathbf{p}_{2},\mathbf{q}_{2})+\psi^{M_{t}}_{\gamma_{3}}(\mathbf{p}_{3},\mathbf{q}_{3}).
\end{eqnarray}
Now we rewrite this equation in the selected coordinate system as:
\begin{eqnarray}\label{eq.27}
\Psi^{M_{t}}_{\gamma}(\mathbf{p},\mathbf{q})
&=&e^{-i[(m_{s_{2}}+m_{s_{3}})\varphi_{p}+(m_{s_{1}}-M_{t})\varphi_{q}]}\,\,\,^{\pm}\psi^{M_{t}}_{\gamma}(p,x_{p},\cos\varphi_{pq},x_{q},q)\nonumber \\
&&+\,e^{-i[(m_{s_{3}}+m_{s_{1}})\varphi_{p_{2}}+(m_{s_{2}}-M_{t})\varphi_{q_{2}}]}
\,\,\,^{\pm}\psi^{M_{t}}_{\gamma_{2}}(p_{2},x_{p_{2}},\cos\varphi_{p_{2}q_{2}},x_{q_{2}},q_{2})\nonumber \\
&&+\,e^{-i[(m_{s_{1}}+m_{s_{2}})\varphi_{p_{3}}+(m_{s_{3}}-M_{t})\varphi_{q_{3}}]}\,\,\,^{\pm}\psi^{M_{t}}_{\gamma_{3}}(p_{3},x_{p_{3}},\cos\varphi_{p_{3}q_{3}},x_{q_{3}},q_{3}).
\end{eqnarray}
By considering:
\begin{eqnarray}\label{eq.64}
&&\varphi_{p_{2}}=\varphi_{q}+\varphi_{p_{2}q},
\,\,\,\,\varphi_{q_{2}}=\varphi_{q}+\varphi_{q_{2}q},\nonumber\\
&&\varphi_{p_{3}}=\varphi_{q}+\varphi_{p_{3}q},
\,\,\,\,\varphi_{q_{3}}=\varphi_{q}+\varphi_{q_{3}q},\quad\quad
\end{eqnarray}
Eq.~(\ref{eq.27}) can be written as:
\begin{eqnarray}\label{eq.66}
&&\,^{\pm}\Psi^{M_{t}}_{\gamma}(p,x_{p},\cos\varphi_{pq},x_{q},q)\nonumber\\&=&
\,^{\pm}\psi^{M_{t}}_{\gamma}(p,x_{p},\cos\varphi_{pq},x_{q},q)+\,e^{i(m_{s_{2}}+m_{s_{3}})\varphi_{pq}}\nonumber \\
&&\times\biggl\{e^{-i[(m_{s_{3}}+m_{s_{1}})\varphi_{p_{2}q}+(m_{s_{2}}-M_{t})\varphi_{q_{2}q}]}\,\,\,^{\pm}\psi^{M_{t}}_{\gamma_{2}}(p_{2},x_{p_{2}},\cos\varphi_{p_{2}q_{2}},x_{q_{2}},q_{2})\nonumber \\
&&+e^{-i[(m_{s_{1}}+m_{s_{2}})\varphi_{p_{3}q}+(m_{s_{3}}-M_{t})\varphi_{q_{3}q}]}\,\,\,^{\pm}\psi^{M_{t}}_{\gamma_{3}}(p_{3},x_{p_{3}},\cos\varphi_{p_{3}q_{3}},x_{q_{3}},q_{3})\biggl\}.\quad\quad
\end{eqnarray}
where:
\begin{eqnarray}\label{eq.65}
p_{2}&=&|-\frac{1}{2}\textbf{p}-\frac{3}{4}\textbf{q}|=\frac{1}{2}\sqrt{p^{2}+\frac{9}{4}q^{2}+3pqy_{pq}},\nonumber\\
p_{3}&=&|-\frac{1}{2}\textbf{p}+\frac{3}{4}\textbf{q}|=\frac{1}{2}\sqrt{p^{2}+\frac{9}{4}q^{2}-3pqy_{pq}},\nonumber\\
q_{2}&=&|\textbf{p}-\frac{1}{2}\textbf{q}|=\sqrt{p^{2}+\frac{1}{4}q^{2}-pqy_{pq}},\nonumber\\q_{3}&=&|-\textbf{p}-\frac{1}{2}\textbf{q}|=\sqrt{p^{2}+\frac{1}{4}q^{2}+pqy_{pq}},\nonumber\\
x_{p_{2}}&=&\hat{\textbf{p}}_{2}\cdot\hat{\textbf{z}}=\frac{-\frac{1}{2}px_{p}-\frac{3}{4}qx_{q}}{p_{2}},\nonumber\\
x_{p_{3}}&=&\hat{\textbf{p}}_{3}\cdot\hat{\textbf{z}}=\frac{-\frac{1}{2}px_{p}+\frac{3}{4}qx_{q}}{p_{3}},\nonumber\\
x_{q_{2}}&=&\hat{\textbf{q}}_{2}\cdot\hat{\textbf{z}}=\frac{px_{p}-\frac{1}{2}qx_{q}}{q_{2}},\nonumber
\\
x_{q_{3}}&=&\hat{\textbf{q}}_{3}\cdot\hat{\textbf{z}}=\frac{-px_{p}-\frac{1}{2}qx_{q}}{q_{3}},\nonumber
\\\label{eq.65}\cos\varphi_{p_{2}q_{2}}&=&\frac{\hat{\textbf{p}}_{2}\cdot\hat{\textbf{q}}_{2}-(\hat{\textbf{p}}_{2}\cdot\hat{\textbf{z}})(\hat{\textbf{q}}_{2}\cdot\hat{\textbf{z}})}{\sqrt{1-(\hat{\textbf{p}}_{2}\cdot\hat{\textbf{z}})^{2}}\sqrt{1-(\hat{\textbf{q}}_{2}\cdot\hat{\textbf{z}})^{2}}}=\frac{\frac{-\frac{1}{2}p^{2}+\frac{3}{8}q^{2}-\frac{1}{2}pqy_{pq}}{p_{2}q_{2}}-x_{p_{2}}x_{q_{2}}}{\sqrt{1-x_{p_{2}}^{2}}\sqrt{1-x_{q_{2}}^{2}}},\nonumber
\\\cos\varphi_{p_{3}q_{3}}&=&\frac{\hat{\textbf{p}}_{3}\cdot\hat{\textbf{q}}_{3}-(\hat{\textbf{p}}_{3}\cdot\hat{\textbf{z}})(\hat{\textbf{q}}_{3}\cdot\hat{\textbf{z}})}{\sqrt{1-(\hat{\textbf{p}}_{3}\cdot\hat{\textbf{z}})^{2}}\sqrt{1-(\hat{\textbf{q}}_{3}\cdot\hat{\textbf{z}})^{2}}}=\frac{\frac{\frac{1}{2}p^{2}-\frac{3}{8}q^{2}-\frac{1}{2}pqy_{pq}}{p_{2}q_{2}}-x_{p_{3}}x_{q_{3}}}{\sqrt{1-x_{p_{3}}^{2}}\sqrt{1-x_{q_{3}}^{2}}},\nonumber
\nonumber\\
\cos\varphi_{p_{2}q}&=&\frac{\hat{\textbf{p}}_{2}\cdot\hat{\textbf{q}}-(\hat{\textbf{p}}_{2}\cdot\hat{\textbf{z}})(\hat{\textbf{q}}\cdot\hat{\textbf{z}})}{\sqrt{1-(\hat{\textbf{p}}_{2}\cdot\hat{\textbf{z}})^{2}}\sqrt{1-(\hat{\textbf{q}}\cdot\hat{\textbf{z}})^{2}}}=\frac{\frac{-\frac{1}{2}py_{pq}-\frac{3}{4}q}{p_{2}}-x_{p_{2}}x_{q}}{\sqrt{1-x_{p_{2}}^{2}}\sqrt{1-x_{q}^{2}}},\nonumber
\\\nonumber\cos\varphi_{p_{3}q}&=&\frac{\hat{\textbf{p}}_{3}\cdot\hat{\textbf{q}}-(\hat{\textbf{p}}_{3}\cdot\hat{\textbf{z}})(\hat{\textbf{q}}\cdot\hat{\textbf{z}})}{\sqrt{1-(\hat{\textbf{p}}_{3}\cdot\hat{\textbf{z}})^{2}}\sqrt{1-(\hat{\textbf{q}}\cdot\hat{\textbf{z}})^{2}}}=\frac{\frac{-\frac{1}{2}py_{pq}+\frac{3}{4}q}{p_{3}}-x_{p_{3}}x_{q}}{\sqrt{1-x_{p_{3}}^{2}}\sqrt{1-x_{q}^{2}}},\nonumber\\
\cos\varphi_{q_{2}q}&=&\frac{\hat{\textbf{q}}_{2}\cdot\hat{\textbf{q}}-(\hat{\textbf{q}}_{2}\cdot\hat{\textbf{z}})(\hat{\textbf{q}}\cdot\hat{\textbf{z}})}{\sqrt{1-(\hat{\textbf{q}}_{2}\cdot\hat{\textbf{z}})^{2}}\sqrt{1-(\hat{\textbf{q}}\cdot\hat{\textbf{z}})^{2}}}=\frac{\frac{py_{pq}-\frac{1}{2}q}{q_{2}}-x_{q_{2}}x_{q}}{\sqrt{1-x_{q_{2}}^{2}}\sqrt{1-x_{q}^{2}}},\nonumber\\
\cos\varphi_{q_{3}q}&=&\frac{\hat{\textbf{q}}_{3}\cdot\hat{\textbf{q}}-(\hat{\textbf{q}}_{3}\cdot\hat{\textbf{z}})(\hat{\textbf{q}}\cdot\hat{\textbf{z}})}{\sqrt{1-(\hat{\textbf{q}}_{3}\cdot\hat{\textbf{z}})^{3}}\sqrt{1-(\hat{\textbf{q}}\cdot\hat{\textbf{z}})^{2}}}=\frac{\frac{-py_{pq}-\frac{1}{2}q}{q_{3}}-x_{q_{3}}x_{q}}{\sqrt{1-x_{q_{3}}^{2}}\sqrt{1-x_{q}^{2}}},
\end{eqnarray}
The labels $\pm$ are related to the signs of $\sin\varphi_{pq}$,
$\sin\varphi_{p_{2}q_{2}}$ and $\sin\varphi_{p_{3}q_{3}}$ which are
determined in appendix D.

\section{Summary and Outlook}\label{sec:summary}
We extend the recently developed formalism for a new treatment of
the Nd scattering in three dimensions for the 3N bound state [13].
We propose a new representation of the 3D Faddeev equation for the
3N bound state including the spin and isospin degrees of freedom in
the momentum space. This formalism is based on 3N free basis state.
This work provides the necessary formalism for the calculation of
the 3N bound state observables which is under preparation.

\appendix

\renewcommand{\theequation}{A.\arabic{equation}}
\setcounter{equation}{0}  

\section*{Appendix A. Anti-symmetrized NN $\textbf{t}$-matrix and its helicity
representation} \label{app:tmatrix}

In our formulation, we need the matrix elements of the
anti-symmetrized NN $t$-matrix. We connect these matrix elements to
the coresponding ones in the momentum-helicity representation. The
antisymmetrized momentum-helicity basis state which is parity
eigenstate is given by~\cite{IFachruddin}:
\begin{eqnarray} \label{momentum-helicity basis}
|\textbf{p};\textbf{\^{p}}S_{23}\lambda;t_{23}\rangle^{\pi a}
&=&\frac{1}{\sqrt{2}}(1-P_{23})|\textbf{p};\textbf{\^{p}}S_{23}\lambda\rangle_{\pi}\,|t_{23}\rangle\nonumber
\\&=&\frac{1}{\sqrt{2}}(1-\eta_{\pi}(-)^{S_{23}+t_{23}})|\textbf{p};\textbf{\^{p}}S_{23}\lambda\rangle_{\pi}\,|t_{23}\rangle,
\end{eqnarray}
Here $S_{23}$ is the total spin, $\lambda$ is the spin projection
along relative momentum of two nucleons, $t_{23}$ is the total
isospin and $|t_{23}\rangle\equiv|t_{23}\tau\rangle$ is the total
isospin state of the two nucleons. $\tau$ is the isospin projection
along its quantization axis which reveals the total electric charge
of system. For simplicity $\tau$ is suppressed since electric charge
is conserved. In Eq. (\ref{momentum-helicity basis}) $P_{23}$ is the
permutation operator which exchanges the two nucleons labels in all
spaces i.e. momentum, spin and isospin spaces, and
$|\textbf{p};\textbf{\^{p}}S_{23}\lambda\rangle_{\pi}$ is parity
eigenstate  which is given by:
\begin{eqnarray}
|\textbf{p};\textbf{\^{p}}S_{23}\lambda\rangle_{\pi}=\frac{1}{\sqrt{2}}(1+\eta_{\pi}P_{\pi})|\textbf{p};
\textbf{\^{p}}S_{23}\lambda\rangle,
\end{eqnarray}
where $P_{\pi}$ is the parity operator, $\eta_{\pi}=\pm1$ are the
parity eigenvalues and
$|\textbf{p};\textbf{\^{p}}S_{23}\lambda\rangle$ is
momentum-helicity state. The Anti-symmetrized two-body $t$-matrix is
given by~\cite{Fachruddin-PRC68}:
\begin{eqnarray}
t_{a}\,_{m_{s_{2}}m_{s_{3}}m_{t_{2}}m_{t_{3}}}^{m'_{s_{2}}m'_{s_{3}}m'_{t_{2}}m'_{t_{3}}}(\mathbf{p},\textbf{p}';\epsilon)&=&\frac{1}{4}
\, \delta _{(m_{t_{2}}
+m_{t_{3}}),(m'_{t_{2}}+m'_{t_{3}})}\,e^{-i(\lambda _{0}\varphi _p
-\lambda _{0}'\varphi _{p' })}\nonumber
\\* &&\times
\, \sum _{ S_{23} t_{23}\pi}\bigl( 1-\eta _{\pi
}(-)^{S_{23}+t_{23}}\bigl) \nonumber\\* &&\times
C(\frac{1}{2}\frac{1}{2}t_{23}; m_{t_{2}} m_{t_{3}} )
\,C(\frac{1}{2}\frac{1}{2}t_{23}; m'_{t_{2}} m'_{t_{3}} )
\nonumber\\* &&\times C(\frac{1}{2}\frac{1}{2}S_{23}; m_{s_{2}}
m_{s_{3}}\lambda_{0} ) \,C(\frac{1}{2}\frac{1}{2}S_{23}; m'_{s_{2}}
m'_{s_{3}}\lambda_{0}' ) \nonumber\\* &&\times \sum _{\lambda
\lambda '}d^{S_{23}}_{\lambda _{0}\lambda
}(x_p)\,d^{S_{23}}_{\lambda _{0}'\lambda '}(x _{p' })\,t_{\lambda
\lambda '}^{\pi S_{23}t_{23}}({{\bf p},{\bf p}' };\epsilon),
\label{eq.t_a-t_helicity}
\end{eqnarray}
where based on momentum-helicity basis states the two-body
$t$-matrix is defined as:
\begin{eqnarray}
t_{\lambda \lambda'}^{\pi S_{23}t_{23}}({{\bf p},{\bf p}'
};\epsilon) \equiv \,^{\pi a}\langle {\bf p};\hat{{\bf
p}}S_{23}\lambda ;t_{23}|t(\epsilon)|{\bf p}';\hat{{\bf
p}}'S_{23}\lambda';t_{23}\rangle^{\pi a},\nonumber\\
\end{eqnarray}
These two-body $t$-matrix elements are connected to the solutions of
Lippmann-Schwinger equation as follow:
\begin{eqnarray}
t_{\lambda \lambda '}^{\pi S_{23}t_{23}}({{\bf p}, \, {\bf p}'
};\epsilon) &=& \frac{\sum ^{S_{23}}_{N=-S_{23}}e^{iN\varphi_{pp'}
}\,d^{S_{23}}_{N\lambda }(x_{p} )\,d^{S_{23}}_{N\lambda '} (x_{p'}
)}{d^{S_{23}}_{\lambda '\lambda }(y_{pp'})} t_{\lambda \lambda
'}^{\pi S_{23}t_{23}}(p, p',y_{pp'};\epsilon),
\end{eqnarray}
where:
\begin{eqnarray}
y_{pp'} =  x_{p}x_{p'}+\sqrt{1-x_{p}^{2}}\sqrt{1-x_{p'}^{2}}
\cos\varphi_{pp'}.
\end{eqnarray}
It should be mentioned that the fully off-shell NN $t$-matrix
$t_{\lambda \lambda '}^{\pi S_{23}t_{23}}(p,p',y_{pp'} ;\epsilon)$,
obeys a set of coupled Lippmann-Schwinger equations which are solved
numerically in Ref.~\cite{IFachruddin}. Finally eq.~(A.3) can be
written as:
\begin{eqnarray}
t_{a}\,_{m_{s_{2}}m_{s_{3}}m_{t_{2}}m_{t_{3}}}^{m'_{s_{2}}m'_{s_{3}}m'_{t_{2}}m'_{t_{3}}}(\mathbf{p},\textbf{p}';\epsilon)&=&
e^{-i[(m_{s_{2}}+m_{s_{3}})\varphi _p
-(m'_{s_{2}}+m'_{s_{3}})\varphi _{p' }]} \nonumber\\&&\times
t^{\pm}_{a}\,_{m_{s_{2}}m_{s_{3}}m_{t_{2}}m_{t_{3}}}^{m'_{s_{2}}m'_{s_{3}}m'_{t_{2}}m'_{t_{3}}}(p,x_{p},\cos
x_{pp'},x_{p'},p',y_{pp'};\epsilon),
\end{eqnarray}
where the labels $\pm$ are related to the sign of
$\sin\varphi_{pp'}$ which is determined as:
\begin{eqnarray}\label{eq.34}
\sin\varphi_{pp'}=\pm\sqrt{1-\cos^{2}\varphi_{pp'}}.
\end{eqnarray}
we consider positive sign for $\varphi_{pp'}\in[0,\pi]$ and negative
sign for $\varphi_{pp'}\in[\pi,2\pi]$.

\renewcommand{\theequation}{B.\arabic{equation}}
\setcounter{equation}{0}  

\section*{Appendix B. Azimuthal dependency of the 3N wave function} \label{app:t matrix2}

We introduce the 3N momentum-helicity basis state as:
\begin{eqnarray}
|\textbf{p};\textbf{\^{p}}S_{23}\lambda,\textbf{q};\textbf{\^{q}}S_{1}\mathrm{\Lambda}\rangle=|\textbf{p};\textbf{\^{p}}S_{23}\lambda\rangle|\textbf{q};\textbf{\^{q}}S_{1}\mathrm{\Lambda}\rangle,
\end{eqnarray}
where:
\begin{eqnarray}
\textbf{S}_{23}\cdot\hat{\textbf{p}}|\textbf{\^{p}}S_{23}\lambda\rangle=\lambda|\textbf{\^{p}}S_{23}\lambda\rangle,\quad\quad\quad\textbf{S}_{1}\cdot\hat{\textbf{q}}|\textbf{\^{q}}S_{1}\mathrm{\Lambda}\rangle=\mathrm{\Lambda}|\textbf{\^{q}}S_{1}\mathrm{\Lambda}\rangle.
\end{eqnarray}
Thus Faddeev component of the 3N wave function can be written as:
\begin{eqnarray}
\psi^{M_{t}}_{\gamma}(\textbf{p},\textbf{q})&=&\sum_{S_{23}\lambda
S_{1}\mathrm{\Lambda}}\langle
\mathbf{p}\mathbf{q}\gamma|\textbf{p};\textbf{\^{p}}S_{23}\lambda,\textbf{q};\textbf{\^{q}}S_{1}\mathrm{\Lambda}\rangle\langle\textbf{p};\textbf{\^{p}}S_{23}\lambda,\textbf{q};\textbf{\^{q}}S_{1}\mathrm{\Lambda}|\psi^{M_{t}}\rangle,
\end{eqnarray}
with considering:
\begin{eqnarray}
|\textbf{\^{p}}S_{23}\lambda\rangle&=&R_{S}(\textbf{\^{p}})|\textbf{\^{z}}S_{23}\lambda\rangle=e^{-iS_{23}^{z}\varphi_{p}}\,e^{-iS_{23}^{y}\theta_{p}}|\textbf{\^{z}}S_{23}\lambda\rangle,\quad\\|\textbf{\^{q}}S_{1}\mathrm{\Lambda}\rangle&=&R_{S}(\textbf{\^{q}})|\textbf{\^{z}}S_{1}\mathrm{\Lambda}\rangle=e^{-iS_{1}^{z}\varphi_{q}}\,e^{-iS_{1}^{y}\theta_{q}}|\textbf{\^{z}}S_{1}\mathrm{\Lambda}\rangle,
\end{eqnarray}
We have written:
\begin{eqnarray}
\langle\mathbf{p}\mathbf{q}\gamma|\textbf{p};\textbf{\^{p}}S_{23}\lambda,\textbf{q};\textbf{\^{q}}S_{1}\mathrm{\Lambda}\rangle&=&\langle\mathbf{p}\mathbf{q}\gamma|R_{S}(\textbf{\^{p}})R_{S}(\textbf{\^{q}})|\textbf{p};\textbf{\^{z}}S_{23}\lambda,\textbf{q};\textbf{\^{z}}S_{1}\mathrm{\Lambda}\rangle\nonumber\\&=&\langle\mathbf{p}\mathbf{q}\gamma|e^{-iS^{z}_{23}\varphi_{p}}\,e^{-iS^{y}_{23}\theta_{p}}\,e^{-iS^{z}_{1}\varphi_{q}}\,e^{-iS^{y}_{1}\theta_{q}}|\textbf{p};\textbf{\^{z}}S_{23}\lambda,\textbf{q};\textbf{\^{z}}S_{1}\mathrm{\Lambda}\rangle\nonumber\\&=&e^{-im_{s_{1}}\varphi_{q}}\,e^{-i(m_{s_{2}}+m_{s_{3}})\varphi_{p}}\langle\mathbf{p}\mathbf{q}\gamma|e^{-iS^{y}_{23}\theta_{p}}\,e^{-iS^{y}_{1}\theta_{q}}|\textbf{p};\textbf{\^{z}}S_{23}\lambda,\textbf{q};\textbf{\^{z}}S_{1}\mathrm{\Lambda}\rangle.\nonumber\\
\end{eqnarray}
Also with considering:
\begin{eqnarray}
|\textbf{p};\textbf{\^{p}}S_{23}\lambda\rangle&=&R_{J_{p}}(\textbf{\^{p}})|p\,\textbf{\^{z}};\textbf{\^{z}}S_{23}\lambda\rangle=e^{-i(L_{p}^{z}+S_{23}^{z})\varphi_{p}}\,e^{-i(L_{p}^{y}+S_{23}^{y})\theta_{p}}|p\,\textbf{\^{z}};\textbf{\^{z}}S_{23}\lambda\rangle,\\|\textbf{q};\textbf{\^{q}}S_{1}\mathrm{\Lambda}\rangle&=&R_{J_{q}}(\textbf{\^{q}})|q\,\textbf{\^{z}};\textbf{\^{z}}S_{1}\mathrm{\Lambda}\rangle=e^{-i(L_{q}^{z}+S_{1}^{z})\varphi_{q}}\,e^{-i(L_{q}^{y}+S_{1}^{y})\theta_{q}}|q\,\textbf{\^{z}};\textbf{\^{z}}S_{1}\mathrm{\Lambda}\rangle,
\end{eqnarray}
We have written:
\begin{eqnarray}
&&\langle\textbf{p};\textbf{\^{p}}S_{23}\lambda,\textbf{q};\textbf{\^{q}}S_{1}\Lambda|\psi^{M_{t}}\rangle\nonumber\\&=&\langle
p\textbf{\^{z}};\textbf{\^{z}}S_{23}\lambda,q\textbf{\^{z}};\textbf{\^{z}}S_{1}\Lambda|R_{J_{p}}^{-1}(\textbf{\^{p}})R_{J_{q}}^{-1}(\textbf{\^{q}})|\psi^{M_{t}}\rangle\nonumber\\&=&\langle
p\,\hat{\textbf{z}};\textbf{\^{z}}S_{23}\lambda,q\hat{\textbf{z}};\textbf{\^{z}}S_{1}\Lambda|e^{i(L^{y}_{p}+S^{y}_{23})\theta_{p}}\,e^{i(L^{z}_{p}+S^{z}_{23})\varphi_{p}}
\,e^{i(L^{y}_{q}+S^{y}_{1})\theta_{q}}\,e^{i(L^{z}_{q}+S^{z}_{1})\varphi_{q}}|\psi^{M_{t}}\rangle\nonumber\\&=&\langle
p\,\hat{\textbf{z}};\textbf{\^{z}}S_{23}\lambda,q\hat{\textbf{z}};\textbf{\^{z}}S_{1}\Lambda|e^{i(L^{y}_{p}+S^{y}_{23})\theta_{p}}\,e^{i(L^{z}_{p}+S^{z}_{23})\varphi_{pq}}
\,e^{i(L^{z}_{p}+S^{z}_{23})\varphi_{q}}\,e^{i(L^{y}_{q}+S^{y}_{1})\theta_{q}}\,e^{i(L^{z}_{q}+S^{z}_{1})\varphi_{q}}|\psi^{M_{t}}\rangle\nonumber\\&=&e^{iM_{t}\varphi_{q}}\langle
p\,\hat{\textbf{z}};\textbf{\^{z}}S_{23}\lambda,q\hat{\textbf{z}};\textbf{\^{z}}S_{1}\Lambda|e^{i(L^{y}_{p}+S^{y}_{23})\theta_{p}}\,e^{i(L^{z}_{p}+S^{z}_{23})\varphi_{pq}}
\,e^{i(L^{y}_{q}+S^{y}_{1})\theta_{q}}|\psi^{M_{t}}\rangle.
\end{eqnarray}
Consequently Eq.~(B.3) can be rewritten as:
\begin{eqnarray}
\psi^{M_{t}}_{\gamma}(\textbf{p},\textbf{q})&=&e^{-i[(m_{s_{2}}+m_{s_{3}})\varphi_{p}+(m_{s_{1}}-M_{t})\varphi_{q}]}\nonumber\\&&\times\sum_{S_{23}\lambda
S_{1}\mathrm{\Lambda}}\langle\mathbf{p}\mathbf{q}\gamma|e^{-iS^{y}_{23}\theta_{p}}\,e^{-iS^{y}_{1}\theta_{q}}|\textbf{p};\textbf{\^{z}}S_{23}\lambda,\textbf{q};\textbf{\^{z}}S_{1}\mathrm{\Lambda}\rangle\nonumber\\&&\times\langle
p\,\hat{\textbf{z}};\textbf{\^{z}}S_{23}\lambda,q\hat{\textbf{z}};\textbf{\^{z}}S_{1}\mathrm{\Lambda}|e^{i(L^{y}_{p}+S^{y}_{23})\theta_{p}}\,e^{i(L^{z}_{p}+S^{z}_{23})\varphi_{pq}}
\,e^{i(L^{y}_{q}+S^{y}_{1})\theta_{q}}|\psi^{M_{t}}\rangle.
\end{eqnarray}
Finally this equation can be written as:
\begin{eqnarray}
\psi^{M_{t}}_{\gamma}(\textbf{p},\textbf{q})&\equiv&e^{-i[(m_{s_{2}}+m_{s_{3}})\varphi_{p}+(m_{s_{1}}-M_{t})\varphi_{q}]}\,\,\,^{\pm}\psi^{M_{t}}_{\gamma}(p,x_{p},\cos\varphi_{
pq},x_{q},q)
\end{eqnarray}

\renewcommand{\theequation}{C.\arabic{equation}}
\setcounter{equation}{0}  

\section*{Appendix C. Parity and time reversal invariance of  the total 3N wave function} \label{app:t matrix2}
In this section we discuss about properties of the total wave
function under the parity and time reversal invariance. Parity
invariance would mean:
\begin{eqnarray}
\langle\textbf{p}\textbf{q}\gamma|\mathrm{\Psi}^{M_{t}}\rangle&=&\langle\textbf{p}\textbf{q}\gamma|P^{-1}_{\pi}P_{\pi}|\mathrm{\Psi}^{M_{t}}\rangle=\langle-\textbf{p},-\textbf{q}\gamma|P_{\pi}|\mathrm{\Psi}^{M_{t}}\rangle=\langle-\textbf{p},-\textbf{q}\gamma|\mathrm{\Psi}^{M_{t}}\rangle\nonumber\\&=&\langle-\textbf{p},-\textbf{q}\gamma|\psi^{M_{t}}\rangle
+\,\langle-\textbf{p}_{2},-\textbf{q}_{2}\gamma_{2}|\psi^{M_{t}}\rangle
+\,\langle-\textbf{p}_{3},-\textbf{q}_{3}\gamma_{3}|\psi^{M_{t}}\rangle,\quad\quad\quad
\end{eqnarray}
where we have used
$P_{\pi}|\mathrm{\Psi}^{M_{t}}\rangle=|\mathrm{\Psi}^{M_{t}}\rangle$
for the 3N total wave function. Eq. (C.1) leads to:
\begin{eqnarray}
\langle\textbf{p},\textbf{q}\gamma|\psi^{M_{t}}\rangle&=&\langle-\textbf{p},-\textbf{q}\gamma|\psi^{M_{t}}\rangle,
\nonumber\\\langle\textbf{p}_{2},\textbf{q}_{2}\gamma_{2}|\psi^{M_{t}}\rangle&=&\langle-\textbf{p}_{2},-\textbf{q}_{2}\gamma_{2}|\psi^{M_{t}}\rangle,
\nonumber\\\langle\textbf{p}_{3},\textbf{q}_{3}\gamma_{3}|\psi^{M_{t}}\rangle&=&\langle-\textbf{p}_{3},-\textbf{q}_{3}\gamma_{3}|\psi^{M_{t}}\rangle.\quad\quad\quad
\end{eqnarray}
So we have:
\begin{eqnarray}
\,^{\pm}\psi_{\gamma}^{M_{t}}(p,x_{p},\cos\varphi_{pq},x_{q},q)&=&e^{-i(M_{s}-M_{t})\pi}\,\,^{\pm}\psi_{\gamma}^{M_{t}}(p,-x_{p},\cos\varphi_{pq},-x_{q},q),
\\\,^{\pm}\mathrm{\Psi}_{\gamma}^{M_{t}}(p,x_{p},\cos\varphi_{pq},x_{q},q)&=&e^{-i(M_{s}-M_{t})\pi}\,^{\pm}\mathrm{\Psi}_{\gamma}^{M_{t}}(p,-x_{p},\cos\varphi_{pq},-x_{q},q),\quad\quad\quad\quad
\end{eqnarray}
where $M_{s}=m_{s_{1}}+m_{s_{2}}+m_{s_{3}}$. Time reversal
invariance of the total wave function can be written as~\cite{Ref2}:
\begin{eqnarray}
\langle\textbf{p}\textbf{q}\gamma|\mathrm{\Psi}^{M_{t}}\rangle&=&\langle\textbf{p}\textbf{q}\gamma|T^{-1}T|\mathrm{\Psi}^{M_{t}}\rangle=i^{2M_{s}}\langle-\textbf{p},-\textbf{q},-\gamma|T|\mathrm{\Psi}^{M_{t}}\rangle=i^{2(M_{s}+M_{t})}\langle-\textbf{p},-\textbf{q},-\gamma|\mathrm{\Psi}^{-M_{t}}\rangle.\nonumber\\
\end{eqnarray}
Considering parity and time reversal invariance lead to:
\begin{eqnarray}
\langle\textbf{p}\textbf{q}\gamma|\mathrm{\Psi}^{M_{t}}\rangle&=&i^{2(M_{s}+M_{t})}\langle\textbf{p}\textbf{q},-\gamma|\mathrm{\Psi}^{-M_{t}}\rangle\nonumber\\
&=&i^{2(M_{s}+M_{t})}\biggl\{\langle\textbf{p}\textbf{q},-\gamma|\psi^{-M_{t}}\rangle
+\,\langle\textbf{p}_{2}\textbf{q}_{2},-\gamma_{2}|\psi^{-M_{t}}\rangle+\langle\textbf{p}_{3}\textbf{q}_{3},-\gamma_{3}|\psi^{-M_{t}}\rangle\biggl\}.\quad\quad
\end{eqnarray}
So we have:
\begin{eqnarray}
&&\,^{\pm}\psi_{\gamma}^{M_{t}}(p,x_{p},\cos\varphi_{pq},x_{q},q)=i^{2(M_{s}+M_{t})\pi}\,\,^{\pm}\psi_{-\gamma}^{-M_{t}}(p,x_{p},\cos\varphi_{pq},x_{q},q),
\quad\quad\\&&\,^{\pm}\mathrm{\Psi}_{\gamma}^{M_{t}}(p,x_{p},\cos\varphi_{pq},x_{q},q)=i^{2(M_{s}+M_{t})\pi}\,^{\pm}\mathrm{\Psi}_{-\gamma}^{-M_{t}}(p,x_{p},\cos\varphi_{pq},x_{q},q).
\end{eqnarray}

\renewcommand{\theequation}{D.\arabic{equation}}
\setcounter{equation}{0}  

\section*{Appendix D. The $\varphi'$-integration}

According to Eq.~(\ref{eq.20}) the $\varphi'$-integration for fixed
$p$, $q$, $x_{p}$, $x_{q}$, $\cos\varphi_{pq}$ and $q'$ can be
written as:
\begin{eqnarray}
I(\varphi_{p},\varphi_{q})&=&\int_{0}^{2\pi}d\varphi'\,e^{im_{1}(\varphi_{q}-\varphi')}\,e^{+im_{2}\varphi_{\pi
q'}}\,e^{-im_{3}\varphi_{\pi'q'}}\nonumber\\&&\times
\,A^{\pm}[\cos(\varphi_{q}-\varphi'),\cos(\varphi_{p}-\varphi'),\cos
\varphi_{pq}] B^{\pm}[\cos(\varphi_{q}-\varphi')],
\end{eqnarray}
where the $A^{\pm}$ and $B^{\pm}$ are known functions determined by
$t^{\pm}_{a}$ and$\,^{\pm}\psi$ respectively. As we know the
exponential functions $e^{+im_{2}\varphi_{\pi q'}}$ and
$e^{-im_{3}\varphi_{\pi'q'}}$ are functions of $\cos\varphi_{\pi
q'}$ and $\cos\varphi_{\pi'q'}$ by considering their sine functions
as:
\begin{eqnarray}
\sin\varphi_{\pi q'}=\pm\sqrt{1-\cos\varphi_{\pi
q'}^{2}},\quad\quad\sin\varphi_{\pi'q'}=\pm\sqrt{1-\cos\varphi_{\pi'q'}^{2}}.
\end{eqnarray}
Also the cosine functions $\cos\varphi_{\pi q'}$ and
$\cos\varphi_{\pi'q'}$ are function of $\varphi_{q}-\varphi'$.
Substituting $\varphi''=\varphi'-\varphi_{q}$ leads to:
\begin{eqnarray}
I(\varphi_{p},\varphi_{q})&=&\int_{0}^{2\pi}d\varphi''\,e^{-im_{1}\varphi''}\,e^{+im_{2}\varphi_{\pi
q'}}\,e^{-im_{3}\varphi_{\pi'q'}}
A^{\pm}[\cos\varphi'',\cos(\varphi_{pq}-\varphi''),\cos\varphi_{pq}]B^{\pm}[\cos\varphi'']\nonumber\\&\equiv&
I^{\pm}(\cos\varphi_{pq}),
\end{eqnarray}
where:
\begin{eqnarray}
&&\cos(\varphi_{pq}-\varphi'')=\cos\varphi_{pq}\cos\varphi''+\sin\varphi_{pq}\sin\varphi'',\quad\quad
\end{eqnarray}
and the labels of $I^{\pm}(\cos\varphi_{pq})$ are depends on the
sign of $\sin\varphi_{pq}$. It is clear that the angles
$\varphi_{\pi q'}$ and $\varphi_{\pi'q'}$ are belong to the interval
$[-\pi,0]$ when $\varphi''$ vary in the interval $[0,\pi]$ and they
are belong to the interval $[0,\pi]$ when $\varphi''$ vary in the
interval $[\pi,2\pi]$. Furthermore since the labels of $B^{\pm}$ are
depend on the sign of $\sin\varphi_{\pi'q'}$, thus for
$\varphi''\in[0,\pi]$ and $\varphi''\in[\pi,2\pi]$ we can choose
negative and positive labels respectively. Consequently the integral
$I^{\pm}(\cos\varphi_{pq})$ can be decomposed as:
\begin{eqnarray}
&&I^{\pm}(\cos\varphi_{pq})\nonumber\\&=&\int_{0}^{\pi}d\varphi''\,e^{-im_{1}\varphi''}\,e^{-im_{2}|\varphi_{\pi
q'}|}\,e^{+im_{3}|\varphi_{\pi'q'}|}
A^{\pm}[\cos\varphi'',\cos(\varphi_{pq}-\varphi''),\cos\varphi_{pq}]B^{-}[\cos\varphi'']\nonumber\\&&+\int_{\pi}^{2\pi}d\varphi''\,e^{-im_{1}\varphi''}\,e^{+im_{2}\varphi_{\pi
q'}}\,e^{-im_{3}\varphi_{\pi'q'}}
A^{\pm}[\cos\varphi'',\cos(\varphi_{pq}-\varphi''),\cos\varphi_{pq}]B^{+}[\cos\varphi''].\nonumber\\
\end{eqnarray}
Now we discuss about the labels of $A^{\pm}$. As we know the labels
of $A^{\pm}$ are related to the sign of $\sin\varphi_{p\pi}$. We can
write $\varphi_{p\pi}=\varphi_{pq}-\varphi_{\pi q}$, and then we
have:
\begin{eqnarray}
\sin\varphi_{p\pi}=\sin\varphi_{pq}\cos\varphi_{\pi
q}-\cos\varphi_{pq}\sin\varphi_{\pi q},
\end{eqnarray}
where:
\begin{eqnarray}
\cos\varphi_{\pi
q}&=&\frac{\hat{\boldsymbol\pi}\cdot\hat{\textbf{q}}-(\hat{\boldsymbol\pi}\cdot\hat{\textbf{z}})(\hat{\textbf{q}}\cdot\hat{\textbf{z}})}{\sqrt{1-(\hat{\boldsymbol\pi}\cdot\hat{\textbf{z}})^{2}}\sqrt{1-(\hat{\textbf{q}}\cdot\hat{\textbf{z}})^{2}}}=\frac{y_{\pi
q}-x_{\pi}x_{q}}{\sqrt{1-x_{\pi}}\sqrt{1-x_{q}^{2}}},\nonumber\\y_{\pi
q}&=&\frac{\frac{1}{2}q+q'y_{qq'}}{\pi}.
\end{eqnarray}
It is clear that the angle $\varphi_{\pi q}$ is belong to the
interval $[0,\pi]$ when $\varphi''$ vary in the interval $[0,\pi]$
and is belong to the interval $[\pi,2\pi]$ when $\varphi''$ vary in
the interval $[\pi,2\pi]$. Thus depending on various intervals of
variables $\varphi_{pq}$ and $\varphi''$,
 we can choose the positive or negative sign for
$\sin\varphi_{pq}$ and $\sin\varphi_{\pi q}$, and then we can
calculate $\sin\varphi_{p\pi}$ from Eq~(B.6). Consequently for
$\sin\varphi_{p\pi}\in[0,1]$ and $\sin\varphi_{p\pi}\in[-1,0]$ we
can consider positive and negative signs of $A^{\pm}$ respectively.
Substituting $\varphi'''=2\pi-\varphi''$, in the second integral of
Eq.~(B.5) yields:
\begin{eqnarray}
&&\int_{0}^{\pi}d\varphi'''\,e^{+im_{1}\varphi'''}\,e^{+im_{2}\varphi_{\pi
q'}}\,e^{-im_{3}\varphi_{\pi'q'}}
A^{\pm}[\cos\varphi''',\cos(\varphi_{pq}+\varphi'''),\cos\varphi_{pq}]\,B^{+}[\cos\varphi'''],\quad\quad\,\,
\end{eqnarray}
Therefore Eq.~(B.5) can be rewritten:
\begin{eqnarray}
I^{\pm}(\cos\varphi_{pq})&=&\int_{0}^{\pi}d\varphi''\,e^{-im_{1}\varphi''}\,e^{-im_{2}|\varphi_{\pi
q'}|}\,e^{+im_{3}|\varphi_{\tilde{\pi}'q'}|}
A^{\pm}[\cos\varphi'',\cos(\varphi_{pq}-\varphi''),\cos\varphi_{pq}]B^{-}[\cos\varphi'']\nonumber\\&&+\int_{0}^{\pi}d\varphi''\,e^{+im_{1}\varphi''}\,e^{+im_{2}\varphi_{\pi
q'}}\,e^{-im_{3}\varphi_{\pi'q'}}
A^{\pm}[\cos\varphi'',\cos(\varphi_{pq}+\varphi''),\cos\varphi_{pq}]B^{+}[\cos\varphi''].\nonumber\\
\end{eqnarray}
According to Eq.~(30) the matrix elements of the total wave
function$\,^{\pm}\mathrm{\Psi}$ can be obtained from the matrix
elements of Faddeev component of the total wave
function$\,^{\pm}\psi$ as follow:
\begin{eqnarray}
\,^{\pm}\mathrm{\Psi_{\gamma}}[\cos\varphi_{pq}]&=&\,^{\pm}\psi_{\gamma}[\cos\varphi_{pq}]+\,e^{+im\varphi_{pq}}\nonumber\\&&\times\biggl\{\,e^{-im'\varphi_{p_{2}q}}\,e^{+im''\varphi_{q_{2}q}}\,^{\pm}\psi_{\gamma_{2}}[\cos\varphi_{p_{2}q_{2}}]+\,
e^{-in'\varphi_{p_{3}q}}\,\,e^{+in''\varphi_{q_{3}q}}\,\,^{\pm}\psi_{\gamma_{3}}[\cos\varphi_{p_{3}q_{3}}]\biggl\},\nonumber\\
\end{eqnarray}
It is clear that for $\varphi_{pq}\in[0,\pi]$, we have:
\begin{eqnarray}
\varphi_{p_{2}q}&\in&[\pi,2\pi],\quad\quad\,\varphi_{q_{2}q}\in[0,\pi],\nonumber\\\varphi_{p_{3}q}&\in&[\pi,2\pi],\quad\quad\,\varphi_{q_{3}q}\in[\pi,2\pi],
\end{eqnarray}
and for $\varphi_{pq}\in[\pi,2\pi]$, we have:
\begin{eqnarray}
\varphi_{p_{2}p}&\in&[0,\pi],\quad\quad\quad\varphi_{q_{2}q}\in[\pi,2\pi],\nonumber\\\varphi_{p_{3}p}&\in&[0,\pi],\:\:\:\,\quad\quad\,\varphi_{q_{3}q}\in[0,\pi].
\end{eqnarray}
Thus Eq.~(D.10) for $\varphi_{pq}\in[0,\pi]$ can be written:
\begin{eqnarray}
\,^{+}\mathrm{\Psi}_{\gamma}[\cos\varphi_{pq}]&=&\,^{+}\psi_{\gamma}[\cos\varphi_{pq}]+\,e^{+im\varphi_{pq}}\nonumber\\&&\times\,\biggl\{e^{+im'\bar{\varphi}_{p_{2}q}}\,e^{+im''\varphi_{q_{2}q}}\,^{\pm}\psi_{\gamma_{2}}[\cos\varphi_{p_{2}q_{2}}]+\,
e^{+in'\bar{\varphi}_{p_{3}q}}\,\,e^{-in''\bar{\varphi}_{q_{3}q}}\,\,^{\pm}\psi_{\gamma_{3}}[\cos\varphi_{p_{3}q_{3}}]\biggl\},\nonumber\\
\end{eqnarray}
and for $\varphi_{pq}\in[\pi,2\pi]$ can be written:
\begin{eqnarray}
\,^{-}\mathrm{\Psi}_{\gamma}[\cos\varphi_{pq}]&=&\,^{-}\psi_{\gamma}[\cos\varphi_{pq}]+\,e^{-im\varphi_{pq}}\nonumber\\&&\times\biggl\{\,e^{-im'\bar{\varphi}_{p_{2}q}}\,e^{-im''\varphi_{q_{2}q}}\,^{\pm}\psi_{\gamma_{2}}[\cos\varphi_{p_{2}q_{2}}]+\,
e^{-in'\bar{\varphi}_{p_{3}q}}\,\,e^{+in''\bar{\varphi}_{q_{3}q}}\,\,^{\pm}\psi_{\gamma_{3}}[\cos\varphi_{p_{3}q_{3}}],\biggl\}\nonumber\\
\end{eqnarray}
where $\bar{\varphi}_{i}=2\pi-\varphi_{i}$. Now we discuss about the
labels of $\,^{\pm}\psi_{\gamma_{2}}$ and
$\,^{\pm}\psi_{\gamma_{3}}$. As we know the labels of
$\,^{\pm}\psi_{\gamma_{2}}$ and $\,^{\pm}\psi_{\gamma_{3}}$ are
related to the signs of $\sin\varphi_{p_{2}q_{2}}$ and
$\sin\varphi_{p_{3}q_{3}}$ respectively. We can write the angles
$\varphi_{p_{2}q_{2}}$ and $\varphi_{p_{3}q_{3}}$ as:
\begin{eqnarray}
\varphi_{p_{2}q_{2}}=\varphi_{p_{2}q}-\varphi_{q_{2}q},\quad\varphi_{p_{3}q_{3}}=\varphi_{p_{3}q}-\varphi_{q_{3}q}.
\end{eqnarray}
Consequently we have:
\begin{eqnarray}
\sin\varphi_{p_{2}q_{2}}&=&\sin\varphi_{p_{2}q}\cos\varphi_{q_{2}q}-\cos\varphi_{p_{2}q}\sin\varphi_{q_{2}q},\nonumber\\\sin\varphi_{p_{3}q_{3}}&=&\sin\varphi_{p_{3}q}\cos\varphi_{q_{3}q}-\cos\varphi_{p_{3}q}\sin\varphi_{q_{3}q}.
\end{eqnarray}
Depending on various intervals of variables $\varphi_{p_{2}q}$,
$\varphi_{q_{2}q}$, $\varphi_{p_{3}q}$ and $\varphi_{q_{3}q}$ we can
choose the positive or negative signs for their sine functions.
Consequently when the calculated $\sin\varphi_{p_{2}q_{2}}$ and
$\sin\varphi_{p_{3}q_{3}}$ from Eq. (D.16) are belong to interval
$[0,1]$ we can use positive sign of $\,^{\pm}\psi_{\gamma_{2}}$ and
$\,^{\pm}\psi_{\gamma_{3}}$ and when they are belong to interval
$[-1,0]$ we can use negative sign of them.

\section{Appendix E. The $x'$-integration}

\renewcommand{\theequation}{E.\arabic{equation}}
\setcounter{equation}{0}  

According to eq.~(\ref{eq.20}) the $x'$-integration carried out as:
\begin{eqnarray}
\,^{\pm}\psi(p,x_{p},\cos\varphi_{pq},x_{q},q)=\int_{-1}^{1}dx'\,
C(x')\,\,^{\pm}\psi(\pi',x_{\pi'},\cos\varphi_{\pi'q'},x',q'),
\end{eqnarray}
where the $C$ is known function determined by $t^{\pm}_{a}$ and
exponential functions. This equation can be rewritten as:
\begin{eqnarray}
\,^{\pm}\psi(p,x_{p},\cos\varphi_{pq},x_{q},q)&=&\int_{-1}^{0}dx'\,
C(x')\,\,^{\pm}\psi(\pi',x_{\pi'},\cos\varphi_{\pi'q'},x',q')\nonumber\\&&+\int_{0}^{1}dx'\,
C(x')\,\,^{\pm}\psi(\pi',x_{\pi'},\cos\varphi_{\pi'q'},x',q')\nonumber\\&=&\int_{0}^{1}dx'\,
C(-x')\,\,^{\pm}\psi(\pi'(-x'),x_{\pi'}(-x'),\cos\varphi_{\pi'q'}(-x'),-x',q')\nonumber\\&&+\int_{0}^{1}dx'\,
C(x')\,\,^{\pm}\psi(\pi',x_{\pi'},\cos\varphi_{\pi'q'},x',q'),
\end{eqnarray}
Finally by considering parity invariance which is described in
appendix C, Eq.~(E.2) can be written:
\begin{eqnarray}
&&\,^{\pm}\psi(p,x_{p},\cos\varphi_{pq},x_{q},q)\nonumber\\&=&\int_{0}^{1}dx'\,
\,\biggl\{\,(-)^{M_{s}+M_{t}}C(-x')\,\,^{\pm}\psi(\pi'(-x'),-x_{\pi'}(-x'),\cos\varphi_{\pi'q'}(-x'),x',q')\nonumber\\&&+\,C(x')\,\,^{\pm}\psi(\pi',x_{\pi'},\cos\varphi_{\pi'q'},x',q')\biggl\}.
\end{eqnarray}

\end{document}